\documentclass[]{spie}  

\usepackage{amsmath,amsfonts,amssymb}
\usepackage{graphicx}
\usepackage[colorlinks=true, allcolors=blue]{hyperref}

\title{Development and task-based evaluation of a scatter-window projection and deep learning-based transmission-less attenuation compensation method for myocardial perfusion SPECT}

\author[a]{Zitong Yu}
\author[a]{Md Ashequr Rahman}
\author[b]{Craig K. Abbey}
\author[c]{Barry A. Siegel}
\author[ac]{Abhinav K. Jha}
\affil[a]{Department of Biomedical Engineering, Washington University in St. Louis, St. Louis, USA}
\affil[b]{Department of Psychological \& Brain Sciences, University of California Santa Barbara, Santa Barbara, USA}
\affil[c]{Mallinckrodt Institute of Radiology, Washington University in St. Louis, St. Louis, USA}

\authorinfo{Further author information: (Send correspondence to Abhinav K. Jha)\\Abhinav K. Jha: E-mail:a.jha@wustl.edu}

\begin{document} 
This manuscript has been accepted to SPIE Medical Imaging, February 19-23, 2023. Please use the following reference when citing the manuscript.

Yu, Z., Rahman, M. A., Abbey, C. K., Siegel, B. A., and Jha, A. K., “Development and task-based evaluation of a scatter-window projection and deep learning-based transmission-less attenuation compensation method for myocardial perfusion SPECT”, Proc. SPIE Medical Imaging, 2023.

\maketitle

\begin{abstract}
Attenuation compensation (AC) is beneficial for visual interpretation tasks in single-photon emission computed tomography (SPECT) myocardial perfusion imaging (MPI). However, traditional AC methods require the availability of a transmission scan, most often a CT scan. This approach has the disadvantage of increased radiation dose, increased scanner costs, and the possibility of inaccurate diagnosis in cases of misregistration between the SPECT and CT images. Further, many SPECT systems do not include a CT component. To address these issues, we developed a Scatter-window projection and deep Learning-based AC (SLAC) method to perform AC without a separate transmission scan. To investigate the clinical efficacy of this method, we then objectively evaluated the performance of this method on the clinical task of detecting perfusion defects on MPI in a retrospective study with anonymized clinical SPECT/CT stress MPI images. The proposed method was compared with CT-based AC (CTAC) and no-AC (NAC) methods. Our results showed that the SLAC method yielded an almost overlapping receiver operating characteristic (ROC) plot and a similar area under the ROC (AUC) to the CTAC method on this task. These results demonstrate the capability of the SLAC method for transmission-less AC in SPECT and motivate further clinical evaluation.  
\end{abstract}


\section{INTRODUCTION}
\label{sec:intro}  

Attenuation of photons is a major image-degrading effect that adversely impacts image quality in single-photon emission computed tomography (SPECT). Multiple studies have shown that attenuation compensation (AC) is beneficial for clinical interpretations of SPECT myocardial perfusion images.\cite{Hutton_2011,garcia2007spect} Conventional AC methods typically require an attenuation map, now most commonly obtained from a separate CT scan. However, these CT-based AC (CTAC) methods have multiple disadvantages, such as increased radiation dose, higher scanner costs, and possible misregistration between the SPECT and CT images potentially leading to inaccurate diagnosis.\cite{li2018mis,sakoshi2018effect,saleki2019influence,yoshikawa2019impact} Further, many SPECT systems often do not have a CT component. For example, SPECT systems in smaller community hospitals and physician offices, as well as mobile SPECT systems facilitating use in remote locations are often SPECT-only. The emerging solid-state-detector-based SPECT systems, which provide higher sensitivity, energy, temporal, and spatial resolution compared to conventional SPECT systems, often do not have CT imaging capability either.\cite{tsuchiya2010basic,suzuki2013high} For these reasons, there is an important need to develop transmission-less AC (Tx-less AC) methods for SPECT.

Given this high significance, multiple Tx-less AC methods have been proposed, including methods that use SPECT emission data to estimate attenuation maps\cite{pan1996segmentation,nunez2009attenuation,zaidi2003determination} and methods that operate on the iterative inversion of the forward mathematical models of SPECT systems.\cite{censor1979new,krol2001algorithm} More recently, deep learning (DL)-based methods have shown significant promise for Tx-less AC.\cite{mcmillan2021ai,hagio2022virtual,yang2021direct,chen2021ct,shi2020deep,chen2022direct} Shi et al. recently reported promising performance of a conditional generative adversarial network for Tx-less AC for myocardial perfusion SPECT (MPS).\cite{shi2020deep} Chen et al. developed strategies for generating attenuation maps using emission data for dedicated cardiac SPECT with small field-of-view and found that their strategies outperformed the method that directly predicts AC images from non-attenuation-corrected images.\cite{chen2022direct} While the performance of these methods is promising, these DL-based methods have typically been evaluated using figures of merit (FoM) that measure the fidelity between the images reconstructed using the DL-based approach with a reference standard, which is typically the image reconstructed with the CT-based AC method. Medical images are acquired for specific clinical tasks. Thus, clinical translation of these Tx-less AC methods requires that they be evaluated in reference to the clinical task.\cite{barrett2015task,barrett1993model,Jha_Objective_AI_eva,jha2022nuclear} However, studies have shown that evaluation using fidelity-based FoMs may not correlate with performance on clinical tasks in myocardial perfusion imaging (MPI).\cite{Yu575,Li_2021,yu2022investigating} Thus, it is crucial to evaluate these methods on the specific clinical tasks for which the images are acquired. 

We have shown that scatter-window data in SPECT contains information to estimate the attenuation distribution.\cite{rahman2020fisher} Based on this premise, we had proposed a DL-based Tx-less AC method for SPECT.\cite{yu2021physics} In this paper, we advance upon this idea to propose a Scatter-window projection and DL-based Tx-less AC (SLAC) method for myocardial perfusion SPECT (MPS) that uses only SPECT emission data in photopeak and scatter windows. We objectively evaluate the method on the clinical task of myocardial defect detection in a retrospective study.

\section{METHODS}
\subsection{Proposed method}
\label{sec:proposed_method}

\begin{figure}[h]
\centerline{\includegraphics[width=0.8\columnwidth]{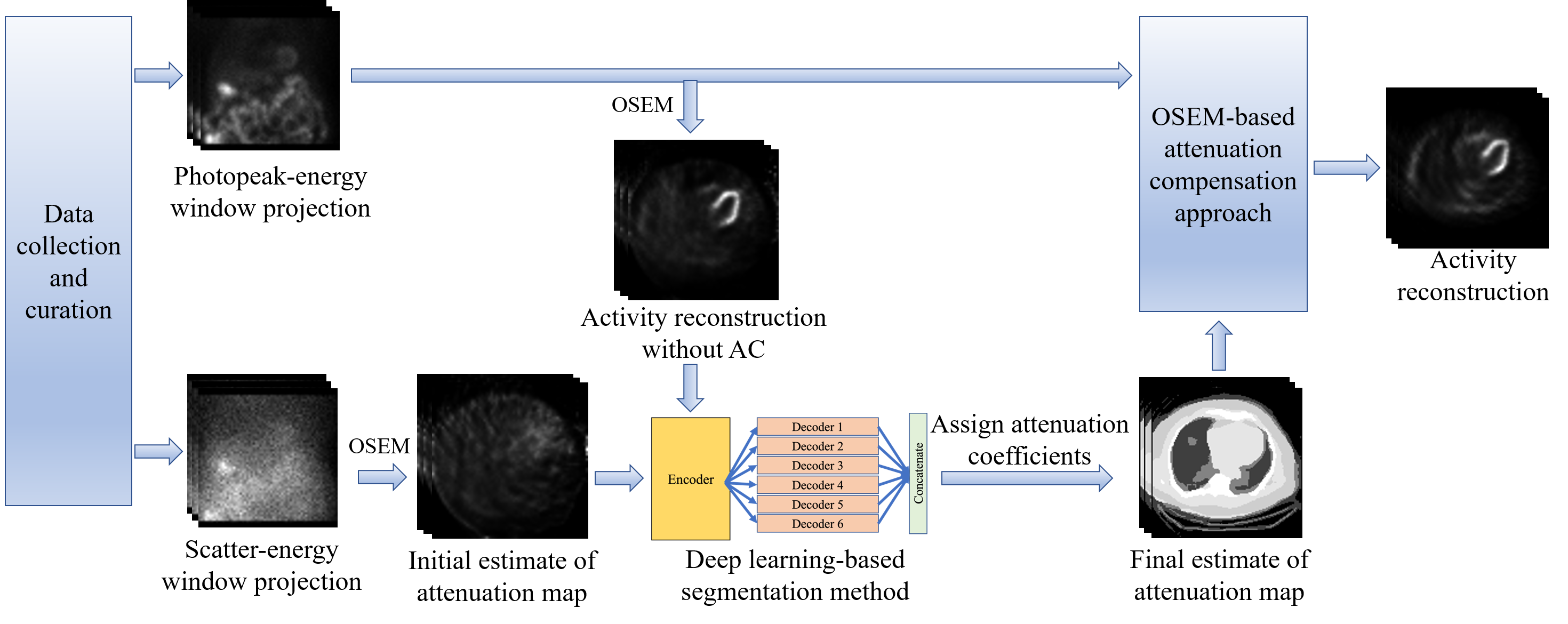}}
\caption{The overall framework of the SLAC method.}
\label{fig:workflow} 
\end{figure}

The overall framework of the SLAC method is shown in Fig.~\ref{fig:workflow}. The probability of scatter at a certain location is proportional to the attenuation distribution at that location. It is expected that a reconstruction of the scatter-window projection would show the contrast between regions with different attenuation coefficients. A previous study has shown a promising performance of a DL-based method that estimates the attenuation map from a reconstruction of scatter-window projection.\cite{shi2020deep} Thus, the scatter-window projection was reconstructed using an ordered-subsets expectation maximization (OSEM)-based approach, yielding an initial estimate of attenuation map.\cite{hudson1994accelerated,merlin2018castor} Then, in this study, we used a DL-based technique to segment the initial estimate of attenuation maps. U-Net-based approaches have shown promise in biomedical image segmentation problems.\cite{ronneberger2015u,leung2020physics,yousefirizi2021toward} Thus, we used a U-Net-based approach, namely, multi-channel input and multi-encoder U-Net (McEUN), to segment the initial estimate of attenuation maps. The McEUN was trained to segment the initial estimate of attenuation maps into six regions, including skin and subcutaneous adipose, muscles and organs, lungs, bones, patient table, and background. The McEUN mainly consists of two components: an encoder with muti-channel input and an assembly of six decoders. To stabilize the network training and leverage salient regions, skip connections with attention gate (AG) were implemented between the output of layers in the encoder and each decoder.\cite{schlemper2019attention} Dropout was applied to prevent overfitting.\cite{srivastava2014dropout} The network was designed to input the whole 3-D image and maximize the amount of information learned in a global sense. The network was trained to minimize the cross entropy between estimated and true segmentation.

\subsection{Evaluation}

We evaluated the SLAC method on the clinical task of detecting cardiac perfusion defects in a retrospective Institutional Review Board-approved evaluation study with anonymized stress SPECT MPI data. We compared the performance of our method with activity maps reconstructed using a CT-based AC (CTAC) method and to images obtained without AC, referred to as no-AC (NAC) approach. We followed the recently proposed Recommendations for EvaLuation of AI in NuClear-medicinE (RELAINCE) guidelines to lend rigor to our evaluation\cite{jha2022nuclear}. The description of the evaluation study consists of four parts, including data collection and curation, network training and method implementation, process to extract task-specific information, and figures of merit used.

\subsubsection{Data collection and curation}
\label{sec:data}
The dataset used in this study consisted of N = 648 anonymized clinical SPECT/CT stress MPI studies scanned between January 2016 and July 2021, with SPECT projection data and CT images along with clinical reports. As per the clinical reports, we categorized patients diagnosed with normal rest and stress myocardial perfusion function as healthy patients, while patients diagnosed with ischemia in a left ventricular wall as diseased patients. MPI scans were acquired on a GE Discovery NM/CT 670 system after the injection of $^{\text{99m}}$Tc-tetrofosmin. SPECT emission data were collected in photopeak (126-154 keV) and scatter windows (114-126 keV). CT images were acquired at 120 kVp on a GE Optima CT 540 system integrated with the GE Discovery NM/CT 670. To avoid misalignment between CT and SPECT scans, CT images were registered to the SPECT space using MIM Maestro (MIM Software Inc, Cleveland, OH). CT-defined attenuation maps were calculated from the CT scans using a bi-linear model.\cite{brown2008investigation}

For evaluating the SLAC method on the clinical task of detecting cardiac defects, knowledge of the existence and location of the defects was needed. The clinical records have limitations in providing this information. To address this issue, we implemented a strategy that introduces synthetic cardiac defects in healthy patient images.\cite{narayanan2001optimization} We designed 27 types of clinically realistic defects with three radial extents, three severities, and at three locations. A summary of defect types is shown in Table~\ref{tab:defect}.

\begin{table}[h]
\caption{Defect parameters}
\label{tab:defect}
\setlength{\tabcolsep}{8pt}
\begin{center}
\begin{tabular}{|c|c|}
\hline
\textbf{Parameter} &  \\
\hline
Extent & 30, 60, and 90 degrees around the left ventricular (LV) wall  \\
\hline
Severity & 10\%, 25\%, and 50\% less activity than the normal myocardium\\
\hline
Location & Anterior, inferior, and lateral LV walls\\
\hline
\end{tabular}
\end{center}
\end{table}

The whole dataset was divided into the training dataset (N = 508) and the testing dataset (N = 140). Because we need the ground truth of defects in the test dataset for our evaluation study, the testing dataset only consisted of patients that were diagnosed as healthy according to the clinical records, and synthetic defects were introduced in 71 of these 140 healthy test patients, referred to as defect-present samples. The remaining 69 healthy test patients were referred to as defect-absent samples. We generated 27×71 = 1917 defect-present samples in both photopeak and scatter energy windows. We also generated 27×69 = 1863 defect-absent samples, although they were identical if from the same patient.

As mentioned in Sec.~\ref{sec:proposed_method}, the scatter-window projection was reconstructed using an OSEM-based reconstruction method without AC, yielding the initial estimate of the attenuation map. Also, the photopeak-window projection was reconstructed using the same strategy, yielding an initial estimate of the activity map. The CT- based attenuation maps for network training were segmented into skin and subcutaneous adipose, muscles and organs, lungs, bones, patient table, and background, using a Markov random field-based method.\cite{zhang2001segmentation} The average attenuation coefficients of each region were calculated and served as the predefined attenuation coefficients.

\subsubsection{Network training and method implementation}

A total of N = 508 samples were used for the network training. The kernel weights of the McEUN were initialized using the Glorot normal initializer.\cite{glorot2010understanding} Biases were initialized to a constant of 0.03. The McEUN was trained to minimize a weighted cross entropy loss between predicted and CT-based segmentations using Adam optimizer.\cite{kingma2014adam} We optimized the weight parameters to yield the best segmentation performance. Five-fold cross-validation was implemented to prevent overfitting. The training and validation were performed using Keras 2.2.4 on two TITAN RTX GPUs with 23 GB memory each.

As mentioned in Sec.~\ref{sec:data}, there were a total of 27×140 = 3780 samples in the test dataset. The trained McEUN yielded segmented masks of initial estimates of attenuation maps. Predefined attenuation coefficients were assigned to each region, yielding the final estimate of attenuation maps. Next, the photopeak-window projections were reconstructed using an OSEM-based reconstruction method, which accounts for attenuation and collimator-detector response.\cite{merlin2018castor} The final estimates of attenuation maps were used for AC. The reconstructed activity images had a size of 64×64×64 with a voxel size of 0.68 cm. Following the clinical protocols, the reconstructed activity images were reoriented into short-axis slices and filtered by a Butterworth filter with an order of 5 and cutoff frequency 0.44 cm$^{-1}$. CTAC-based images and NAC-based images were obtained using the same OSEM-based reconstruction approach and post-processing procedures as used in the SLAC method but with different AC approaches.

\subsubsection{Process to extract task-specific information}

We objectively evaluated the performance of the SLAC method on the task of detecting myocardial perfusion defects in an observer study. While ideally, such evaluation should be performed with human observers, this is time-consuming and tedious. Model observers provide an easy-to-use in silico approach to perform such evaluation and identify methods for evaluation with human observers. Thus, multiple studies use model observers to evaluate imaging systems and methods.\cite{barrett1993model,frey2002application,he2004mathematical,Li_2017} In our evaluation study, the location of the defect was chosen to be in the inferior left ventricular wall in the test dataset. Then, we had a patient population where the defect extent and severity were varying, but the location of the defect was the same in the entire population in the test dataset. Therefore, there were 9×71 = 639 defect-present samples and 9×69 = 621 defect-absent samples in the evaluation study. Previous studies have shown that the channelized Hotelling observer (CHO) with rotationally symmetric frequency channels can emulate human-observer performance on the task of detecting perfusion defects from MPS images in this setting.\cite{wollenweber1999comparison,sankaran2002optimum} Thus, in this study, we used this model observer. We extracted a 32×32 region from the middle 2-D slice of the short-axis images that had the defect centroid at the center and applied the CHO to this region to yield test statistics. These statistics were calculated for each defect-present and -absent image in the test set using a leave-one-out strategy.

\subsubsection{Figures of merit}
The test statistics generated were compared to a threshold to classify the image into the defect-present or defect-absent class. By varying the threshold, the ROC curve was plotted\cite{barrett2013foundations,metz1986roc} using LABROC4 program.\cite{metz1998maximum} The AUC measures the performance of methods on the task of defect detection. A higher AUC indicates better performance. We calculated the AUC with 95\% confidence intervals (CIs) for SLAC, CTAC, and NAC methods. 

To assess the performance of our method using visual fidelity-based criterion, we calculated root mean-square-error (RMSE) and structural similarity index (SSIM) with CIs between images obtained using the SLAC and the CTAC method, as well as between images obtained using the NAC and the CTAC method.

\section{RESULTS}
\subsection{Comparing to other AC methods}

\begin{figure}
\centerline{\includegraphics[width=0.5\columnwidth]{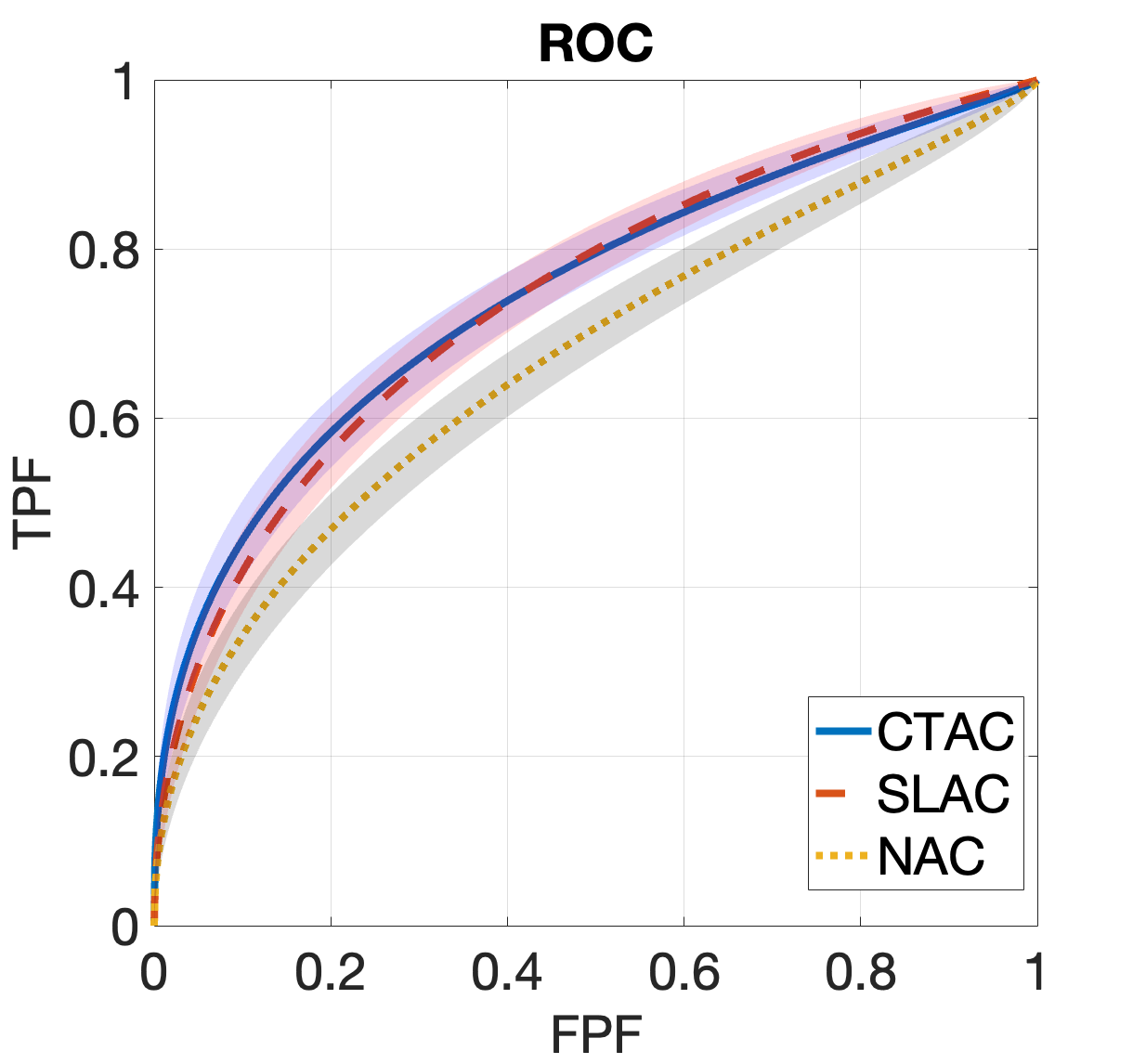}}
\caption{ROC curves obtained by CTAC, SLAC, and NAC methods. Shadows indicate 95\% confidence intervals.}
\label{fig:roc} 
\end{figure}

Fig.~\ref{fig:roc} shows ROC curves obtained by SLAC, CTAC, and NAC methods, along with the corresponding 95\% CIs. The ROC curve obtained by the SLAC method almost overlapped those obtained by the CTAC method and outperformed the NAC method. Fig. 3 shows the AUC values with CIs obtained by three methods. We observed that the AUC obtained by SLAC was similar to that obtained by the CTAC method and significantly outperformed ($p<0.05$) the NAC method.

\begin{figure}
\centerline{\includegraphics[width=0.5\columnwidth]{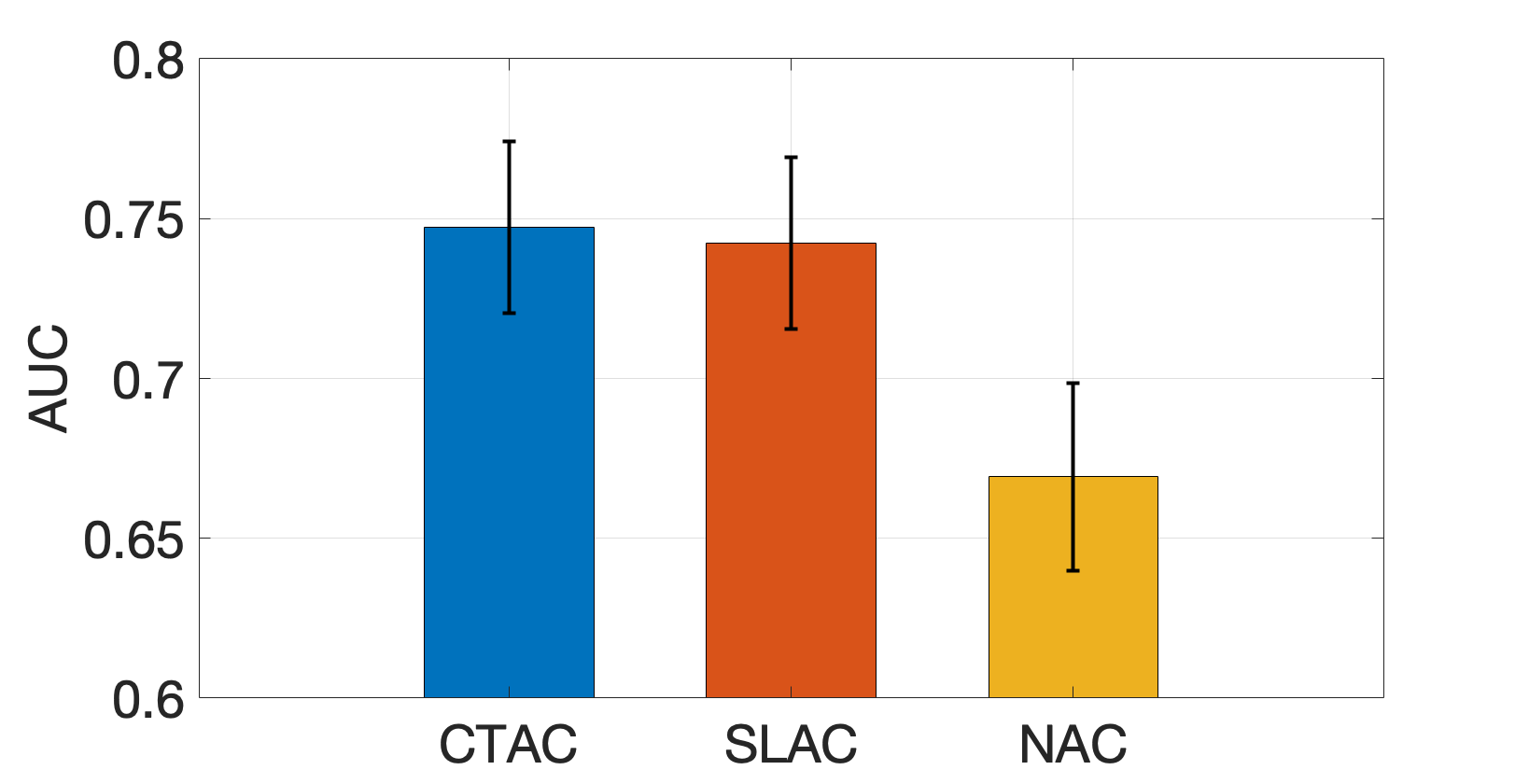}}
\caption{AUC obtained by CTAC, SLAC, and NAC methods.}
\label{fig:auc} 
\end{figure}

\subsection{Representative examples}
Fig. 4 shows examples of SPECT images and corresponding attenuation maps estimated by SLAC, compared with those from CTAC. We found that the attenuation maps obtained using SLAC method were close to those obtained from CT images. Further, the short-axis SPECT images obtained using SLAC methods were similar to those obtained using CTAC methods. Fig. 4.b shows the same defect was shown in CTAC and SLAC-based images.

\begin{figure}
\centerline{\includegraphics[width=0.8\columnwidth]{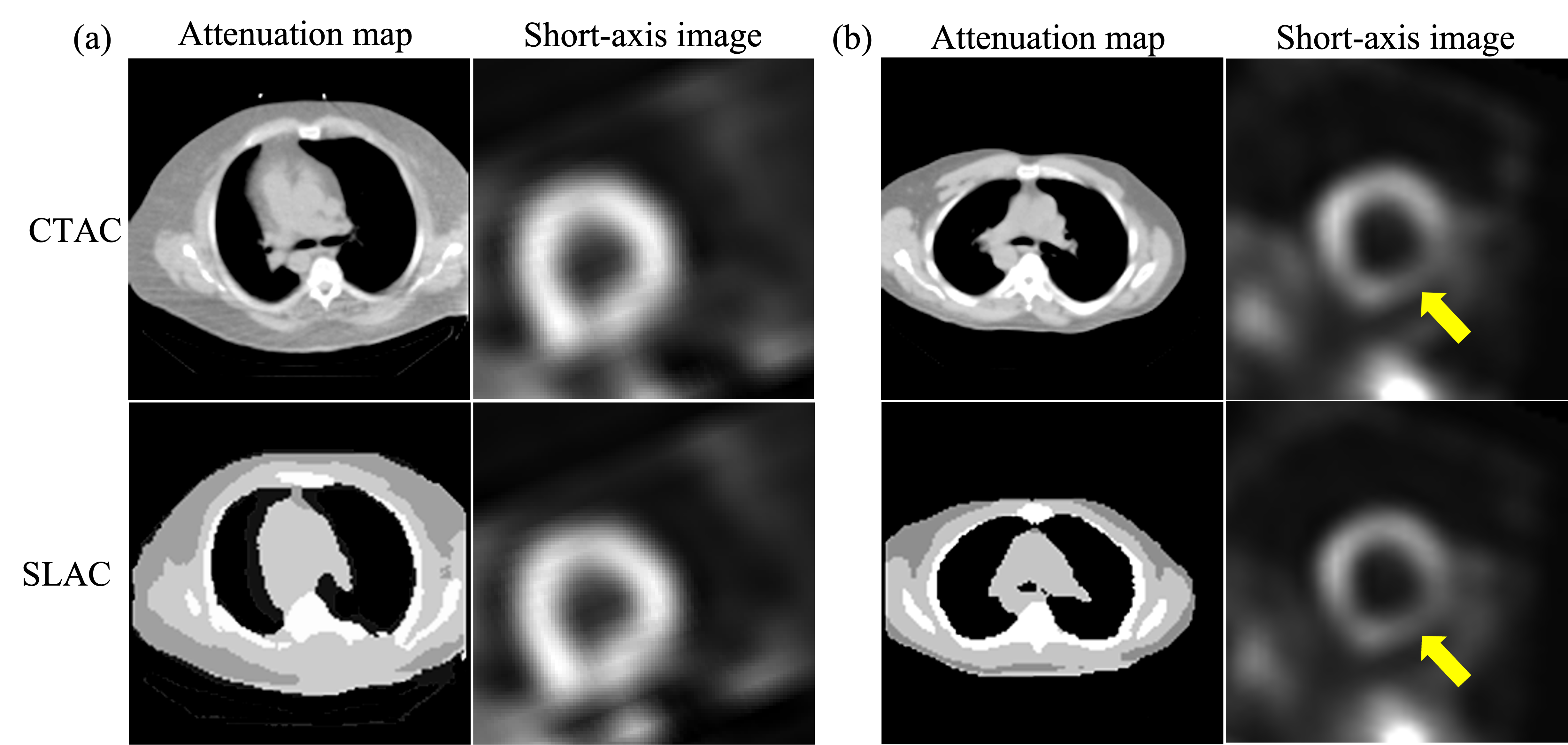}}
\caption{Examples of SPECT images and attenuation maps obtained by SLAC and CTAC method: (a) a defect-absent example, (b) a defect-present example, where the yellow arrow indicates the detect.}
\label{fig:example} 
\end{figure}

\subsection{Evaluation using fidelity-based figures of merit}

Table~\ref{tab:convFoMs} shows RMSE and SSIM between images obtained using the SLAC and the CTAC method, as well as between the NAC and the CTAC method. We found that the SLAC method significantly outperformed the NAC method based on these metrics.

\begin{table}[h]
\caption{Evaluation using fidelity-based FoMs}
\label{tab:convFoMs}
\setlength{\tabcolsep}{8pt}
\begin{center}
\begin{tabular}{|c|c|c|}
\hline
\textbf{Method} & \textbf{RMSE (95\% CIs)} & \textbf{SSIM (95\% CIs)} \\
\hline
SLAC & 0.013 (0.011, 0.014) & 0.96 (0.95,0.97)  \\
\hline
NAC & 0.036 (0.032,0.040) & 0.66 (0.65,0.68)\\
\hline
\end{tabular}
\end{center}
\end{table}

\section{DISCUSSIONS AND CONCLUSIONS}

This study represents a step along the pathway for developing a method for AC in SPECT without requiring a transmission scan. Our results demonstrate that the proposed SLAC method did not just yield activity images that were visually similar to those yielded by the CTAC method (as evaluated using the RMSE and SSIM metrics) (Table~\ref{tab:convFoMs} and Fig.~\ref{fig:example}), but also yielded a similar performance to the CTAC method on the clinical task of detecting perfusion defects from MPI images (Fig.~\ref{fig:auc}). More specifically, the proposed method yielded an almost overlapping ROC and similar AUC to the CTAC method and significantly outperformed the NAC method. Further, evaluation using fidelity-based FoMs showed that the SLAC method significantly outperformed the NAC method (Table~\ref{tab:convFoMs}). These results demonstrate that the SLAC method has the capability of performing AC using only emission data for SPECT.

Our evaluation results motivate evaluation of the SLAC method with data from different scanners and from different institutions to assess for generalizability. Another area of future research is advancing the method to use list-mode data instead of sinogram data to reconstruct the attenuation map. Previous studies have shown that processing data in list-mode format in SPECT can yield improved performance on clinical tasks compared to processing data in binned format.\cite{rahman2021task,rahman2020list,clarkson2020quantifying,jha2015singular,jha2015estimating} More specifically, it has been quantitatively shown that list-mode data contains more information compared to binned data for the task of estimating attenuation coefficients.\cite{rahman2020fisher} Thus, advancing the method to process data in list-mode format directly may lead to even more improved performance.

There are some limitations to this study. First, the performance of the SLAC method is dependent on the quality of training data, including the segmentation of CT images. We used a Markov random field-based method to segment the CT images. However, there are other methods, including DL-based methods, which may yield segmented low-dose CT images with better segmentation performance.\cite{van2013automated,fu2021review} Integrating a more advanced segmentation method is a future direction for improving the performance of the SLAC method. Second, we used a model observer study to evaluate the defect detection performance of the SLAC method. While the observer we used has been shown to mimic human observer performance, ideally this study should be conducted using human observers, such as experienced radiologists. The results from the model observer study motivate the evaluation of the method using human observers. Furthermore, the test dataset only consisted of defects located in the inferior LV wall. A future research direction is to incorporate defects at multiple locations in the test dataset while performing the observer study.

In conclusion, a scatter window projection and deep learning-based transmission-less AC method for SPECT yielded similar performance compared to the standard CTAC method, as evaluated on the clinically relevant task of detecting myocardial perfusion defects with a model observer in a retrospective evaluation study.

\acknowledgments 
 This work was partly supported by National Institute of Biomedical Imaging and Bioengineering of National Institute of Health (NIH) under grant number R21-EB024647, R01-EB031051, and R01EB031962. Support is also acknowledged from the NVIDIA GPU grant. We also thank the Washington University Center for High Performance Computing for providing computational resources. The center is partially funded by NIH grants 1S10RR022984-01A1 and 1S10OD018091-01. We also thank Dr. Craig K. Abbey for helpful discussions on the design of the observer study.

\bibliography{report} 
\bibliographystyle{spiebib} 

\end{document}